\input harvmac
\input epsf
\def\half{{\scriptstyle{\scriptstyle 1\over \scriptstyle 2}}}
\lref\thomas{E.L.~Thomas, D.M.~Anderson, C.S.~Henkee and D.~Hoffman, 
``Periodic Area-Minimizing Surfaces in Block Copolymers'', Nature {\bf 334}
(6183), 598--601 (1988).}

\lref\dgp{See, for instance, P.-G. de Gennes and J. Prost, {\sl The Physics
of Liquid Crystals}, Second Edition, Oxford University Press, New York, (1993).}

\lref\kl{R.D.~Kamien and T.C.~Lubensky, ``Minimal Surfaces, Screw 
Dislocations and Twist Grain Boundaries'', Phys. Rev. Lett. {\bf 82} (14), 
2892--2895 (1999).}

\lref\rama{S.~Ramanujan, {\sl Ramanujan's Notebooks}, (Edited by Bruce
C. Berndt), chapter 2, entry 11, 
Springer-Verlag, New York, (1985).
.}

\lref\bromwich{T.J.I'A.~Bromwich, {\sl An Introduction to the Theory
of Infinite Series, $2^{\rm nd}$ Edition}, Macmillan, London, (1942).}

\lref\Scherk{H.F.~Scherk, ``Bemerkungen \"uber die kleinste Fl\"ache 
innerhalb gegebener Grenzen'', J. r. angew. Math. {\bf 13}, 185--208 (1835); 
see also \nitsche .}   

\lref\nitsche{J.C.C.~Nitsche, {\sl Vorlesungen \"uber Minimalfl\"achen},
Springer-Verlag, Berlin, (1975); J.C.C.~Nitsche, {\sl Lectures on Minimal
Surfaces},
(Translated by J.M.~Feinberg), Cambridge University Press, Cambridge, (1989).}

\lref\barbcher{B.M.~Barbashov and N.A.~Chernikov, 
``Solution of the Two Plane Wave Scattering Problem in a Nonlinear Scalar
Field Theory of the Born-Infeld Type'', J. Exptl. Theoret. Phys. {\bf 51} (2),
658--668 (1966) [Soviet Physics JETP, {\bf 24} (2), 437--442 (1967)]; see also
\whitham.}

\lref\whitham{G.B.~Whitham, {\sl Linear and Nonlinear Waves}, 
Wiley, New York, (1974).}

\centerline{\bf Decomposition of the Height Function of} 
\centerline{\bf Scherk's First Surface}

\centerline{Randall D. Kamien\footnote{$^\dagger$}{\tt
kamien@physics.upenn.edu}}\smallskip
\centerline{\sl Department of Physics and
Astronomy, University of Pennsylvania, Philadelphia,
PA 19104}

\centerline{16 August 2000; revised 29 September 2000}
\smallskip
\noindent{\bf Abstract} We show that Scherk's first surface, a one-parameter
family of solutions to the
minimal surface equation, may be written as a linear superposition 
of other solutions with specific parametric values.\hfill\break

\bigskip

A surface in the Monge representation,
represented by its height function $h[x,y]$, is a minimal surface whenever
\eqn\eminimal{(1+h_y^2)h_{xx} - 2h_xh_yh_{xy} + (1+h_x^2)h_{yy}=0.}
A well-known solution is 
Scherk's first surface \Scherk , shown in Fig. 1:
\eqn\escherk{z=h[x,y;\alpha]\equiv-\sec({1\over 2}\alpha)
\tan^{-1}\left\{{
\tanh\left[{1\over 2}x\sin(\alpha)\right]\over
\tan\left[y\sin({1\over 2}\alpha)\right]}\right\}
}
As $x\rightarrow\pm\infty$, $z=\pm y\tan({1\over 2}\alpha) 
+ (n+\half)\pi\sec({1\over 2}\alpha)$, respectively for integers $n$.  
Thus Scherk's first surface connects two
infinite sets of parallel planes at $x=\pm\infty$, equally spaced by
$\pi$ and rotated by an angle $\alpha$ with respect to each other.  It
has been used as a model for grain boundaries in diblock copolymers and
smectic liquid crystals \refs{\thomas,\kl} where the multi-valued 
height function 
represents the peak of the one-dimensional density modulation of these 
materials.
In the limit that $\alpha\rightarrow 0$, the solution $z=h[x,y;\alpha]$ in
\escherk\ is the height function for another well-known minimal surface, the helicoid:
\eqn\eheli{
\lim_{\alpha\rightarrow 0} 
h[x,y;\alpha]
= -\tan^{-1}\left\{{
{\alpha}x\over
{\alpha}y}\right\}=\tan^{-1}
\left({y\over x}\right) - {\pi\over 2}.}
The helicoid corresponds to a topological defect (specifically a screw dislocation)
in a quiescent layered structure \dgp .  As we shall see, Scherk's first surface
is an infinite superposition of these topological defects.

\vskip20pt
\vbox{
\epsfxsize=5truein
\centerline{\epsfbox{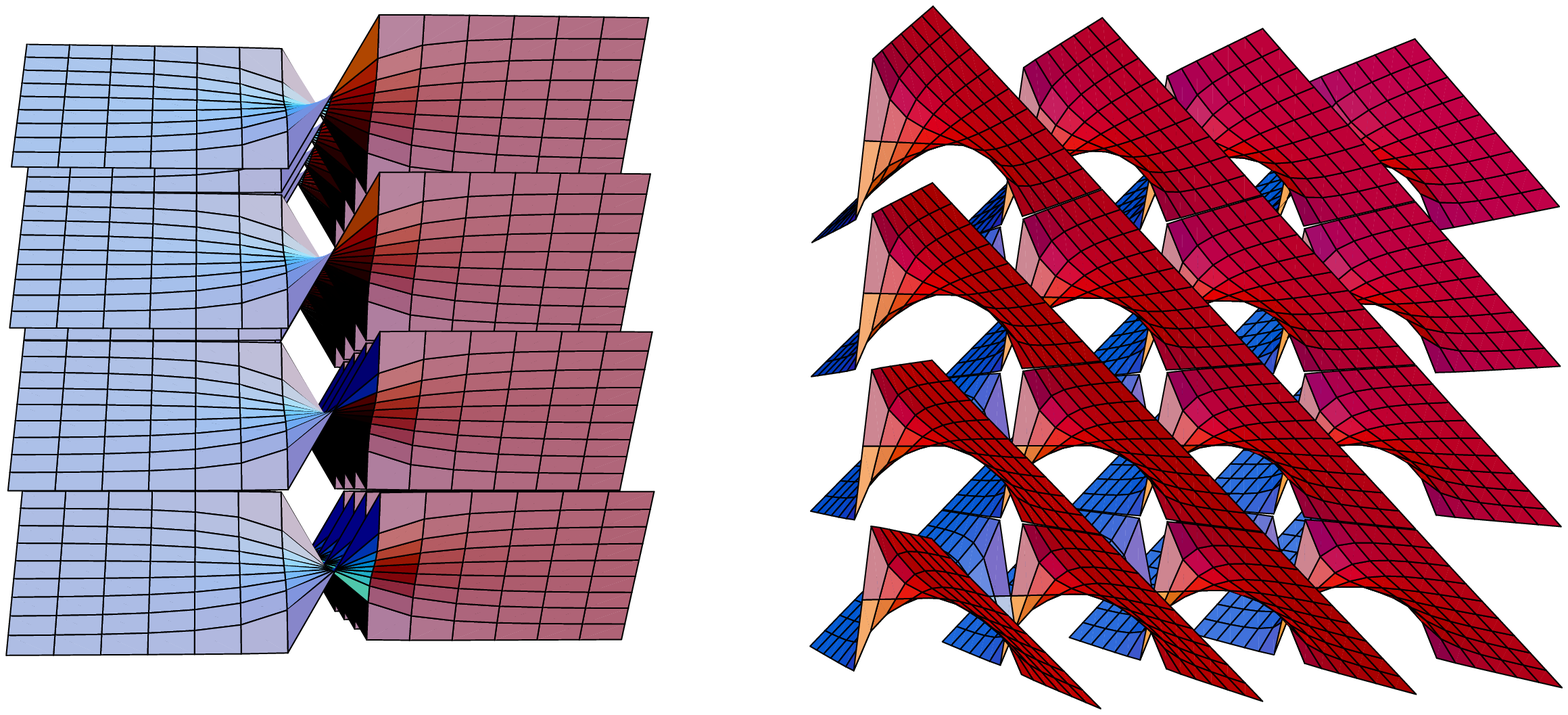}}
\vskip 10pt
\centerline{\vbox{\hsize=5truein \noindent{\bf Figure 1} 
Scherk's First
Surface (two views).  This structure connects together two 
lamellar structures with
different orientations.
}}}
\vskip 10pt

It has long been known that the arctangent function satisfies a
number of remarkable identities \rama .  Employing these we have two
results:

\noindent {\bf Theorem 1}: 
The solution $h[x,y;\alpha]$ of \escherk\ may
be decomposed into an infinite superposition of dilated helicoids:
\eqn\ekl
{h[x,y;\alpha] = \sec(\half\alpha)\sum_{n=-\infty}^\infty
\left\{\tan^{-1}\left({y-n\ell\over x\cos(\half\alpha)}\right)
-{\pi\over 2}\right\}}
where $\sin(\half\alpha)={\pi\over\ell}$ defines $\ell$.  

\noindent {\sl Proof}: This follows from a result of Ramanujan \rama :
\eqn\eram{\tan^{-1}\left[\tanh a\cot b\right] = \sum_{k=-\infty}^\infty
\tan^{-1}\left({a\over b + k\pi}\right)}
by taking $a=\half x\sin(\alpha)$ and $b=y\sin(\half\alpha)$.  
The result \ekl\ may also be derived via the Poisson summation formula 
applied to derivatives of \escherk .  That derivation only proves
the equality of \ekl\ and \eram\ up to an additive constant.  In \kl\ 
an infinite additive constant was neglected in comparison with
\ekl .  However since the height function may be arbitrarily shifted
along the $z$-axis, an additive constant is irrelevant.  It is important
to note that the dilation
preserves the topology, but not the geometry of the helicoids and
so the dilated helicoids are neither true helicoids nor
are they minimal surfaces.  However, 
it is surprising that a pure dilation along $x$ and not a
function of $x$ makes a sum of helicoids into a minimal surface.  

A class of finite decompositions of \escherk\ are also possible:

\noindent {\bf Theorem 2}:  The solution $h[x,y;\alpha]$ of \escherk\ may
be decomposed into a finite superposition of dilated Scherk's first surfaces:
\eqn\efinite{h[x\sec\beta,y;2\beta] = 
{\cos\tilde\beta\over\cos\beta}\sum_{m=0}^{n-1}
h[x\sec\tilde\beta,y+{m\over n}\pi\csc\tilde\beta;2\tilde\beta]}
with $\sin\beta=n\sin\tilde\beta$.  

\noindent{\sl Proof}: 
This follows by noting that
\eqn\etaninv{
\tan^{-1}\left[{\tanh x\over\tan y}\right] = \Im \ln\sin(y+ix)} 
and the identity \bromwich :
\eqn\esin{\sin(nz) = 
2^{n-1}\prod_{m=0}^{n-1} \sin\left(z+{m\pi\over n}\right).}

An alternative proof is based on \eram .  Note that for an
arbitrary positive integer $n$, the sum can be decomposed
into a ``sum of sums'':
\eqnn\esumofsum{$$\eqalign{
\sum_{k=-\infty}^\infty \tan^{-1}\left({a\over b+k\pi}\right) &=
\sum_{m=0}^{n-1} \sum_{p=-\infty}^\infty \tan^{-1}\left(
{a\over b + (np+m)\pi}\right)\cr &= \sum_{m=0}^{n-1}\tan^{-1}\left[
\tanh\left({a\over n}\right)\cot\left({b+m\pi\over n}\right)\right]\cr}$$}
Again, the dilation of the solutions by $\sec\tilde\beta$ preserves the
topology of the ``subboundaries''.  Note, however, that for small $x$,
$h[x,y;\alpha]$ is linear in $x$ and so the factors of $\sec\beta$ multiplying
$x$ will cancel the factors of $\cos\beta$ multiplying $h$.  Thus, in some sense,
\efinite\ extends the small $x$ limit, where true, undistorted linear
superposition holds, to the entire $x$ axis.
 
Note that as the subboundaries are
moved apart $\ell$ necessarily grows, $\alpha$ tends toward $0$, the
factors of cosine and secant tend to unity, and true linear superposition
prevails.  This is reminiscent of
the multi-soliton solutions of the KdV or sine-Gordon
equations -- as the superposed solutions are moved further apart,
undistorted linear superposition prevails \whitham  .  Unlike these integrable
systems, however, in moving the solutions apart we must
change the angle of rotation $\alpha$ which also alters the
form of the subboundaries.  Nonetheless, these decompositions suggest that
$h[x,y;\alpha]$ has greater symmetry than arbitrary solutions of \eminimal .  

\lref\kp{R.D.~Kamien and T.R.~Powers, in preparation (2000).}
The decompositions provided by the two theorems presented here should 
prove useful in studying the stability of grain boundaries in 
layered systems.  Scherk's first surface has been used to model 
such boundaries \thomas .  However it is not clear that
Scherk's first surface minimizes nonlinear elastic energy functionals
such as those considered in \kl .  By using these decompositions it is
possible to study long wavelength perturbations to the boundaries.  For instance,
while a typical approach might consider the stability of the location of
a single topological defect, by utilizing \efinite\ for $n=2$, we can shift
half of the topological defects with respect to the other half and thus
probe long-wavelength deformations of the grain boundary.  This work is
in progress \kp .  

A variational approach to minimizing the energy functional would also be
possible.  For instance, one could deform the functional form of $h[x,y;\alpha]$ via
\eqn\evari{\tilde h[x,y;\alpha,\gamma] = h[{\rm sgn}(x)\vert x\vert^\gamma,y;\alpha],}
calculate the energy as a function of $\gamma$, and then minimize over $\gamma$.  
The decompositions here would allow a much broader class of variational {\sl ansetzen} that
would still be computationally manageable.

Though the remarkable linear superposition properties of
this solution might suggest a connection to the linear
Weierstrass-Enneper representation of minimal surfaces \nitsche, this is,
unfortunately, not the case. 
The Weierstrass-Enneper representation allows
for linear superposition of a minimal surface in parametric form, whereas
\efinite\ is in nonparametric form: we are unaware of any connection
between that representation and our result. 

Whether the hidden symmetry that gives rise to these decompositions can lead
to similar decompositions of other height functions of minimal surfaces 
is an open but interesting question.
It is amusing to note that the Born-Infeld equation:
\eqn\ebi{(1-\phi_t^2)\phi_{xx} + 2\phi_x\phi_y\phi_{xy} 
- (1+\phi_x^2)\phi_{yy}=0}
is related to the minimal surface equation through the Wick rotation
$y=it$.  It is worth noting that $f(x-t)$ and $f(x+t)$ are 
each solutions to \ebi\ for any $f(\cdot)$, and that soliton-like scattering
properties exist for this equation \barbcher .  Unfortunately, upon
analytic continuation these solutions become complex and
are inadmissable as height functions.  Moreover,
the hodograph transformation,
on which the results in \barbcher\ rely, 
cannot be performed when branch-cut singularities are present as in \eheli .

It is a pleasure to acknowledge stimulating conversations with
M.~Cohen, C.~Epstein, G.~Jungman, H.~Karcher, T.~Lubensky, R.~Kohn and T.~Powers.  This work
was supported in part by NSF CAREER Grant DMR97-32963, the Alfred P. Sloan
Foundation and a gift from L.J.~Bernstein.

\listrefs
\bye